# Negative refractive perfect lens vs Spherical geodesic lens. Perfect Imaging comparative analysis.

Juan C. González, Pablo Benítez, Juan C. Miñano, Dejan Grabovičkić

Universidad Politécnica de Madrid, Cedint
Campus de Montegancedo
28223 Madrid, Spain

E-mail: jcgonzalez@cedint.upm.es

**Abstract.** Negative Refractive Lens (NRL) has shown that an optical system can produce images with details below the classic Abbe diffraction limit. This optical system transmits the electric field, emitted by the object surface, towards the image surface producing the same field distribution in both surfaces. In particular, a Dirac delta electric field in the object surface is focused without diffraction limit to the Dirac delta electric field in the image surface. The Maxwell Fish Eye lens (MFE) and the Spherical Geodesic Waveguide (SGW) have been claimed to break the diffraction limit using positive refraction with a different meaning. In these cases, it has been considered the power transmission from a point source to a point receptor, which falls drastically when the receptor is displaced from the focus by a distance much smaller than the wavelength. Although these systems can detect displacements up to $\lambda/3000$, they cannot be compared to the NRL, since the concept of the object and image surface is not established. Here, it is presented an analysis of the SGW with defined object and image surfaces. The results show that a Dirac delta electric field on the object surface produces an image below the refraction limits on the image surface.

## 1. Introduction

Perfect Imaging (PI) stands for the capacity of an optical system to produce images with details below the classic Abbe diffraction limit. In 2000 Pendry [1] proposed a new device built with material of negative dielectric and magnetic constant (left-handed materials [2]) that reaches the theoretical limit of PI. In the last decade PI has been shown experimentally with devices made of left-handed materials [3][4] but unfortunately, high absorption and small (wavelength scale) source-to-image distance are both present in these experiments.

An alternative device for PI has recently been proposed [6][7]: the Maxwell Fish Eye (MFE) lens. Unlike previous PI devices, MFE uses materials with a positive, isotropic refractive index distribution. This device is well known in the framework of Geometrical Optics because it is an Absolute Instrument [8], so every object point has a stigmatic image point.

Leonhardt [6] analyzed Helmholtz wave fields in the MFE lens in two dimensions (2D). These Helmholtz wave fields describe TE-polarized modes in a cylindrical MFE, i.e., modes in which electric field vector points orthogonally to the cross section of the cylinder. Leonhardt found a family of Helmholtz wave fields which have a monopole asymptotic behavior at an object point as well as at its stigmatic image point. Each one of these solutions describes a wave propagating from the object point to the image point. It coincides asymptotically with an outward (monopole) Helmholtz wave at the object point, as generated by a point source, and with an inward (monopole) wave at the image point, as it was sunk by an "infinitely-well localized drain" (which we call a "perfect point drain"). This perfect point drain absorbs the incident wave, with no reflection or scattering. This result has also been confirmed via a different approach [9].

Two sets of experiments have recently been carried out to support the PI capability in the MFE. In the first one, PI with positive refraction has been demonstrated for the very first time at a microwave-frequency ($\lambda$=3 cm) [10][11]. The second set of experiments has been carried out for the near infrared frequency ($\lambda$ = 1.55 µm), but PI with resolution below the diffraction limit was not found [12]. The authors assume that the failure in the experimental demonstration is due to manufacturing flaws in the prototype.

Recently, it has been analyzed PI properties of the Spherical Geodesic Waveguide (SGW) for microwave frequencies [13]. The SGW, a device suggested in [14], is a spherical waveguide filled with a non-magnetic material and isotropic refractive index distribution proportional to 1/$r$ ($\varepsilon = (r_0/r)^2$ and $\mu = 1$), $r$ being the distance to the center of the spheres. Transformation Optics theory [15] proves that the TE-polarized electric modes of the cylindrical MFE are transformed into radial-polarized modes in the SGW, so both have the same imaging properties. When the waveguide thickness is small enough, the variation of the refractive index within the two spherical shells can be ignored resulting in a constant refractive index within the waveguide. The analyzed device is formed by the SGW and two coaxial probes (source and drain) loaded with specific designed impedances. The results show that this device present up to $\lambda$/500 resolution for a discrete number of frequencies, called notch frequencies, that are close to the well known Schumman Resonance frequencies of spherical systems. For other frequencies the system does not present resolution below diffraction limit.

The theoretical result presented by Leonhard [6], the experimental set up described in [11] and the analysis of the SGW [13] show a concept of perfect imaging different from the NRL one presented by Pendry in [1]:

- In the NRL lens, the radiation emitted by the object surface with tangent electric field $\mathbf{E}(x,y,z_1)$ (plane x-y in $z=z_1$) is formed by a complete set of plane waves travelling in z direction. The optical system transmits the plane wave to the image surface (plane x-y in $z=z_2$) generating the same field $\mathbf{E}(x,y,z_1)$= $\mathbf{E}(x,y,z_2)$. In particular, if the field in the object plane is $\mathbf{E}(x,y,z_1)$=$\delta(x,y)\mathbf{y}$ (Dirac delta), the field in the image plane is $\mathbf{E}(x,y,z_2)$=$\delta(x,y)\mathbf{y}$, Fig. 1.
- In the theoretical and experimental results for the MFE and SGW does not exist the concept of object and image surface, but only point source and drain. The electric field has asymptotic behavior at the point source and drain when they are placed in the focal points. In the case of the SGW, it has been shown [13] that a small displacement of the drain (much smaller than the wavelength) from the focal point has the effect of radical change in the field distribution and a drastic fall in the transmitted power from the source to the drain.

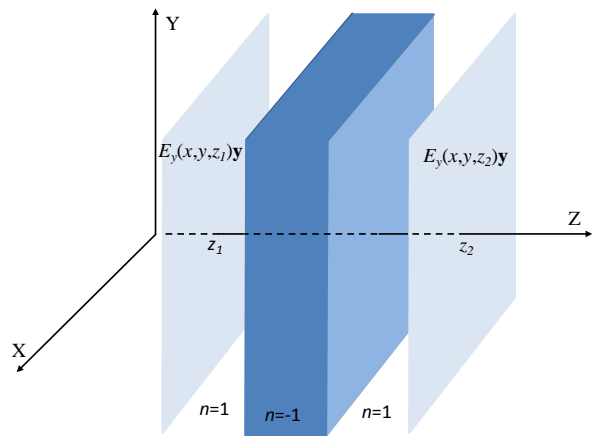

Fig. 1. Pendry lens. The object surface ($z=z_1$) with tangent field $E(x,y,z_1)\mathbf{y}$ radiates power in the direction z. The radiation reaches to the image plane ($z=z_2$) and generates the tangent electric field $E(x,y,z_2)\mathbf{y}$. When the condition $E(x,y,z_1)$= $E(x,y,z_2)$ is fulfilled, the system performs Perfect Imaging.

Here, it is presented an analysis of the SGW based on the concept of object and image surface. The object surface is $\theta=\theta_0$ (equivalent to $z=z_1$ of the NRL) and the image surface is $\theta_I=\pi-\theta_0$ (equivalent to $z=z_2$ of the NRL), (see Fig.3 and Fig. 4). In the object surface, the tangent electric field is $\mathbf{E}(r,\theta_0,\alpha)$, while in the image surface it is $\mathbf{E}(r,\theta_I,\alpha)$. In a perfect imaging system, the fields in the object and image surface have to be equal. Here, it will be shown that, in general, the condition $\mathbf{E}(r,\theta_0,\alpha)= \mathbf{E}(r,\theta_I,\alpha)$ is not fulfilled, except for the limit case $\theta_0 \to 0$ that corresponds to the point source and drain. However, a very interesting result has been found: if the field in the object surface is $\mathbf{E}(r,\theta_0,\alpha)=\delta(\alpha)\mathbf{r}$, then the field in the image surface is $\mathbf{E}(r,\theta_I,\alpha)= A\delta(\pi-\alpha)\mathbf{r}+f(\alpha)\mathbf{r}$, where $f(\alpha)$ is a finite function and $A$ is a constant. This means that although the SGW doesn´t work as a perfect imaging system, it can focus below the Abbe diffraction limit.

Section 2 describes in details the Pendry lens and the SGW, showing the conceptual similarity of the two systems. Modal analysis of the SWG, presented in section 3, shows an important result in relation with perfect imaging: there is a complete set of non-evanescent waves that transmits the information from the object to the image surface. In the classical case of free space, the complete set of waves that transmits the information is composed by evanescent and non-evanescent waves. The negative refraction is necessary to recover the lost information of the evanescent waves. The results are shown in section 4. Discussion and conclusions are presented in section 5.

## 2 Negative refractive lens.

The system described by Pendry in [1] is shown in Fig. 2 (see also Fig. 1):
- The object plane is placed in $z=0$, and the image plane in $z=Z_L$.
- The electric field has only y component and does not depend on y coordinate. The magnetic field has x,z, components and does not depend on y coordinate:

$$\mathbf{E}(x,y,z) = E_y(x,z)\mathbf{y} \qquad \mathbf{H}(x,y,z) = H_x(x,z)\mathbf{x} + H_z(x,z)\mathbf{z} \qquad (1)$$

- The relative dielectric and magnetic constants are as follow:

$$\varepsilon(z)=\begin{cases}1 & 0<z<Z_L/4 \\ -1 & Z_L/4<z<3Z_L/4 \\ 1 & 3Z_L/4<z<Z_L\end{cases} \qquad \mu(z)=\begin{cases}1 & 0<z<Z_L/4 \\ -1 & Z_L/4<z<3Z_L/4 \\ 1 & 3Z_L/4<z<Z_L\end{cases} \qquad (2)$$

- The waves coming from the object plane are not reflected at the image plane ($z=Z_L$). That is, $\varepsilon(z)=1$ and $\mu(z)=1$ for $z > Z_L$. If there is no reflection at the image plane for each mode, then the receptor is called Perfect Drain.
- Fig. 3.a shows the ray trajectory (plane x-z), while Fig.3.b shows the electric and magnetic field vectors.

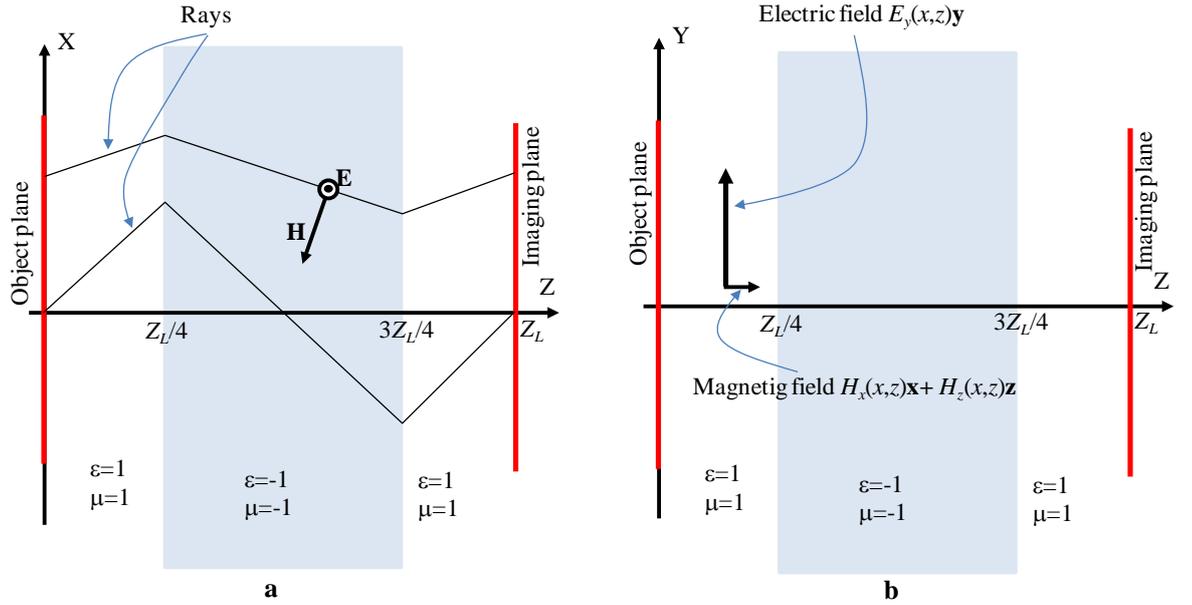

Fig. 2. Structure of the Negative Refractive Lens. a. The ray trajectories in the plane x-z. b. Plane y-z, the electric vector has only y component and does not depend on y coordinate.

- If two ideal metallic planes are placed normal to the axis Y, the fields between the two planes are not affected and have the same value (Fig. 3). The original structure is modified in this way to make the comparative analysis with the spherical lens.

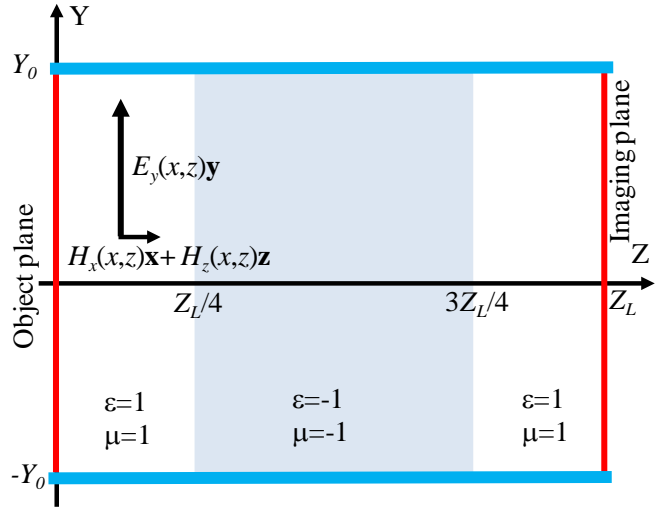

Fig. 3. Modified structure of the negative refraction lens. The two parallel metallic planes do not alter the value of the fields.

The EM fields in the region $0 < z < Z_L/4$, emitted by the object plane, can be expressed as a set of incident waves:

$$\mathbf{E}(x,z) = \left( \int_{-k_0}^{k_0} E_i(k_x) e^{-jk_x x} e^{-jz\sqrt{k_0^2 - k_x^2}} dk_x + \int_{|k_0|<k_x} E_i(k_x) e^{-jk_x x} e^{-z\sqrt{k_x^2 - k_0^2}} dk_x \right) \mathbf{y} \quad (3)$$

The first integral groups the transmitted waves and the second the evanescent ones. In the object plane (z=0) this expression is:

$$\mathbf{E}(x,0) = \int_{-\infty}^{\infty} E_i(k_x) e^{-jk_x x} dk_x \mathbf{y} \qquad (4)$$

Where clearly $E_i(k_x)$ can be obtained as the Inverse Fourier Transform of the tangent field in the object plane.
In the classical optical system with positive refraction, the evanescent waves are not transmitted towards the image plane, thus a non-perfect image of the original is obtained:

$$\mathbf{E}(x,Z_L) = \int_{-k_0}^{k_0} E_i(k_x) e^{-jk_x x} dk_x \qquad (5)$$

For the particular case of Dirac delta distribution:

$$\mathbf{E}(x,0) = \delta(x)\mathbf{y} = \int_{-\infty}^{\infty} e^{-jk_x x} dk_x \mathbf{y} \qquad \mathbf{E}(x,Z_L) = \int_{-k_0}^{k_0} e^{-jk_x x} dk_x \mathbf{y} = 2k_0 \frac{\sin(k_0 x)}{k_0 x} \mathbf{y} \qquad (6)$$

This shows the diffraction limit for two dimensional radiation. The NRL has the property of recovering the evanescent waves in the image plane, thus the tangential electric field is as in Eq. (4).

### 3. Spherical Geodesic Waveguide.

#### 3.1. Structure.

The SGW analyzed here is shown in Fig. 4:
- The structure is formed by two concentric metallic spheres with gradual dielectric constant.
- The object surface is $\theta=\theta_0$ and the image surface is $\theta=\pi-\theta_0$.
- The electric field has only $r$ component and does not depend on $r$ coordinate. The magnetic field has components $\theta$, $\varphi$ and does not depend on $r$ coordinate:

$$\mathbf{E}(r,\theta,\varphi) = E_r(\theta,\varphi)\mathbf{r} \qquad \mathbf{H}(r,\theta,\varphi) = H_\theta(\theta,\varphi)\boldsymbol{\theta} + H_\varphi(\theta,\varphi)\boldsymbol{\varphi} \qquad (7)$$

- The relative dielectric and magnetic constants are as follow:

$$\varepsilon(r,\theta,\varphi) = \varepsilon_0 \left(\frac{R_M}{r}\right)^2 \qquad \mu(r,\theta,\varphi) = \mu_0 \qquad (8)$$

- There is no any reflection of the waves arriving to the image surface $\theta=\pi-\theta_0$. The receptor is considered as a Perfect Drain.

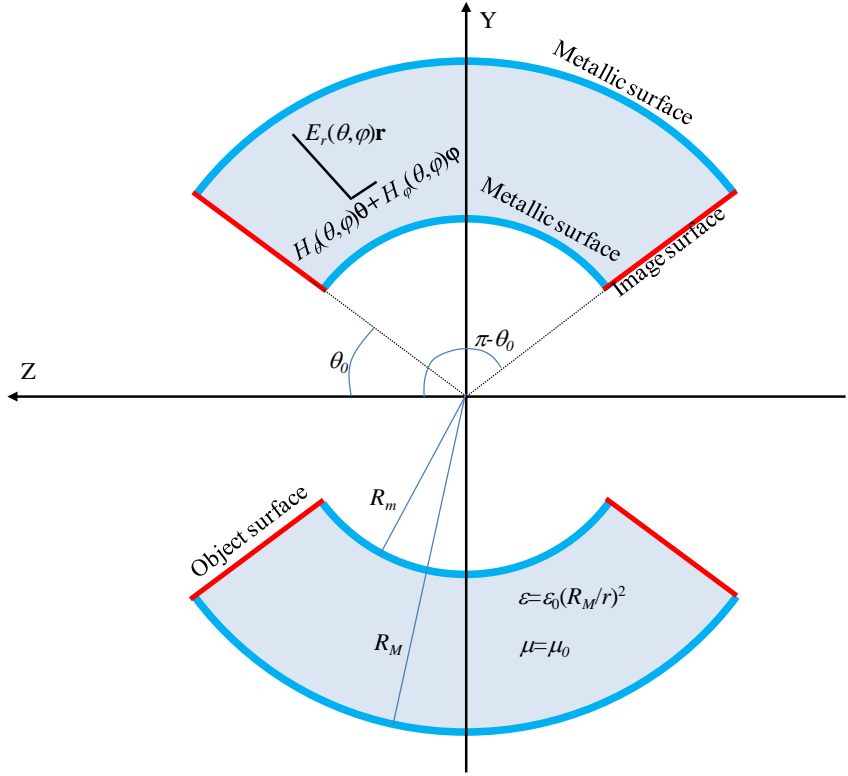

Fig. 4. Structure of the SGW. It is formed by two concentric metallic spheres and filled with dielectric and magnetic constant depending on r. The object surface is $\theta=\theta_0$ and the image surface is $\theta=\pi-\theta_0$.

### 3.2. Modal analysis of the SWG.

The electric and magnetic fields **E** and **H** expressed in Eq. (7) with the EM constants defined in Eq. (8) are given as:

$$\mathbf{E}(\theta,\varphi) = E(\theta)e^{jn\varphi}\mathbf{r}$$
$$\mathbf{H}(\theta,\varphi) = \frac{\nabla \times \mathbf{E}(\theta,\varphi)}{-j\omega\mu_0} = e^{jn\varphi}\left[-\frac{E(\theta)jn}{\sin(\theta)}\boldsymbol{\theta} + \frac{dE(\theta)}{d\theta}\boldsymbol{\varphi}\right]\frac{1}{j\omega\mu_0 r} \quad (9)$$

From the second Maxwell equation:

$$\mathbf{E}(\theta,\varphi) = \frac{\nabla \times \mathbf{H}(\theta,\varphi)}{j\omega\varepsilon}$$
$$\frac{d}{d\theta}\left(\sin(\theta)\frac{dE(\theta)}{d\theta}\right) + E(\theta)\left((k_0 R_M)^2 \sin(\theta) - \frac{n^2}{\sin(\theta)}\right) = 0 \quad (10)$$

The solution is:

$$E(\theta) = A P_\nu^n(\cos(\theta)) + B Q_\nu^n(\cos(\theta))$$
$$\nu(\nu+1) = (k_0 R_M)^2 \quad (11)$$

Where $P_\nu^n(x)$ and $Q_\nu^n(x)$ are the Legendre function of first and second type, $A$ and $B$ integration constant and $n$ an integer constant (for the condition $\mathbf{E}(\theta,0)=\mathbf{E}(\theta,2\pi)$) (10). $E_r(\theta)$

can be expressed as follows:

$$E(\theta) = E_f F_\nu^n(\cos(\theta)) + E_r R_\nu^n(\cos(\theta))$$

$$F_\nu^n(\cos(\theta)) = P_\nu^n(\cos(\theta)) + ja\frac{2}{\pi}Q_\nu^n(\cos(\theta)) \qquad (12)$$

$$R_\nu^n(\cos(\theta)) = P_\nu^n(\cos(\theta)) - ja\frac{2}{\pi}Q_\nu^n(\cos(\theta))$$

Where $F_\nu^n(x)$ and $R_\nu^n(x)$ are called forward and reverse Legendre function [16], $E_f$, $E_r$ are constants and $a$ is a constant with value +1 or -1, which will be explained bellow. The fields **E** and **H** can be developed in modes given by Eq. (9) and (12):

$$\mathbf{E}(\theta,\varphi) = \sum_n \left[ E_{fn} F_\nu^n(\cos(\theta)) + E_{rn} R_\nu^n(\cos(\theta)) \right] e^{jn\varphi} \mathbf{r}$$

$$\mathbf{H}(\theta,\varphi) = \frac{1}{j\omega\mu_0} \sum_n e^{jn\varphi} \left[ \begin{array}{l} \dfrac{-(E_{fn} F_\nu^n(\cos(\theta)) + E_{rn} R_\nu^n(\cos(\theta)))jn}{\sin(\theta)}\boldsymbol{\theta} + \\ \dfrac{d(E_{fn} F_\nu^n(\cos(\theta)) + E_{rn} R_\nu^n(\cos(\theta)))}{d\theta}\boldsymbol{\varphi} \end{array} \right] \qquad (13)$$

### 3.3. Poynting vector and transmitted power in direction $\theta$.

The Poynting vector and the transmitted power through a surface defined by $\theta$=cte, are:

$$\mathbf{S} = \frac{1}{2}\mathrm{Re}[\mathbf{E}\times\mathbf{H}^*] \qquad S_\theta = \frac{1}{2}\mathrm{Re}[\mathbf{E}\times\mathbf{H}^*]\cdot\boldsymbol{\theta}$$

$$P_W(\theta) = \int_{R_{min}}^{R_{max}} \int_0^{2\pi} S_\theta r\sin(\theta)drd\varphi = \frac{(R_{max}^2 - R_{min}^2)}{\omega\mu_0} a \sum_n A(\nu,n)(|E_{fn}|^2 - |E_{rn}|^2) \qquad (14)$$

Where $A$ is a constant that depends on the frequency ($\nu$) and the mode ($n$). This result is obtained from orthogonallity of exponential functions from Eq. (13) and the following property of the Legendre Function [16]:

$$(P_\nu^n \frac{dQ_\nu^n(\cos(\theta))}{d\theta} - Q_\nu^n \frac{dP_\nu^n(\cos(\theta))}{d\theta})\sin(\theta) = cte = A(\nu,n) \qquad (15)$$

Where the constant is real and can be positive or negative depending on the values of $\nu$ and $n$. The results expressed in Eq. (14) show that the fields in Eq. (13) are composed by an orthogonal set of incident and reflected waves. The constant $a$ is 1.0 for $A > 0$ and -1.0 for $A < 0$.
In the object surface (Fig. 4) the electric field is as follows (see Eq. (13)):

$$\mathbf{E}(\theta_0,\varphi) = \sum_n \left[ E_{fn} F_\nu^n(\cos(\theta_0)) + E_{rn} R_\nu^n(\cos(\theta_0)) \right] e^{jn\varphi}\mathbf{r} = \sum_n A_n e^{jn\varphi}\mathbf{r} \qquad (16)$$

The field tangent to the object source can be expressed as a Fourier series. According to the definition of the incident and reflected waves (see Eq. (12)) and the result from Eq. (14) we can express the field as formed only by incident waves:

$$\mathbf{E}(\theta_0,\varphi) = \sum_n A_n e^{jn\varphi} \mathbf{r} = \sum_n E_{fn} F_\nu^n(\cos(\theta_0)) e^{jn\varphi} \mathbf{r} \qquad (17)$$

This equation is equivalent to Eq. (4) obtained in the case of the NRL.

### 3.4. Comparative imaging analysis NRL-SGW.

The same analysis is done for the SGW and the NRL (see Fig. 2 and Fig. 4).
- In the SGW the object and image surfaces are the surfaces $\theta=\theta_0$ and $\theta=\pi-\theta_0$. In the NRL these surfaces are the planes $z=0$ and $z=Z_L$.
- In the SGW the electric field has only $r$ component and does not depend on coordinate $r$. In the NRL the electric field has only y component and does not depend on coordinate y.
- In the SGW the field emitted by the object surface can be expressed as in Eq. (17). The coefficients $E_{fn}$ are obtained from the boundary conditions, $\mathbf{E}(\theta_0,\varphi)$. In the NRL, the field is expressed as in Eq.(4), while the coefficients $E_f(k_x)$ are obtained from the boundary conditions, $\mathbf{E}(y,0)$.
- The image surface is considered as a Perfect Drain, that is, all the power of the incident wave $F_\nu^n$ will be absorbed without reflection or scattering. The same case is considered in [1] for the NRL.

Fig. 5 shows the tangent field distribution in the image plane of the NRL (right-hand side) when the object plane has the tangent field distribution defined as on the left-hand side of Fig. 5. As it is well known, in the case of the NRL, the fields in the object and image planes coincide perfectly. Note that the distance is expressed in wavelengths.

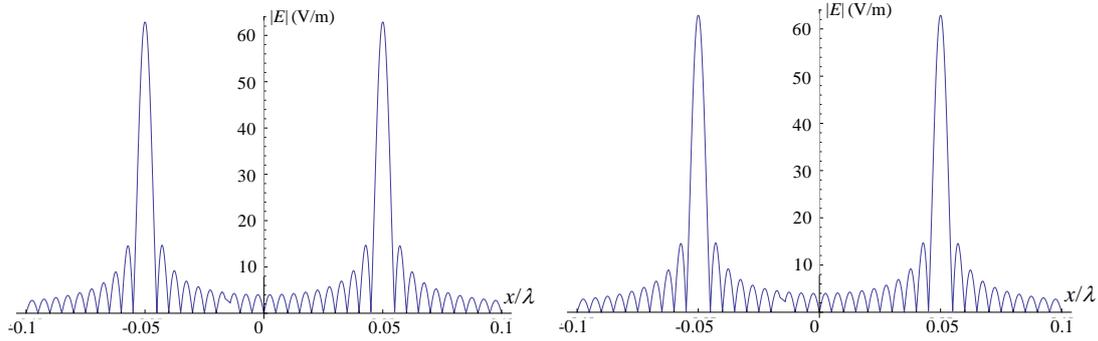

Fig. 5. Modulus of the tangent electric field in the object plane (left) and image plane (right) in the NRL. In the three parts of the system, $0 < z < Z_L/4$, $Z_L/4 < z < 3Z_L/4$ and $3Z_L/4 < z < Z_L$, there are only incident waves. The separation between the two peaks is $\lambda/10$

Fig. 6 shows the tangent field distribution in the image surface of the SGW (right-hand side) when the object surface has the tangent field distribution defined as on the left-hand side of Fig. 6. The field is expressed as a function of the azimutal variable $\varphi$ of the spherical coordinate system. The waves between the object and image surface are the incident waves as defined in Eq.(17). The dimensions of the SGW are: $R_m=10\lambda$, $\theta_0=0.2\pi$. Fig. 6 shows that two quasi-Dirac delta distributions, placed in two points $\varphi_1$ and $\varphi_2$ and separated by $\lambda/10$, are focused in two quasi-Dirac delta distributions, placed in $\varphi_1'=\varphi_1-\pi$ and $\varphi_2'=\varphi_2-\pi$.

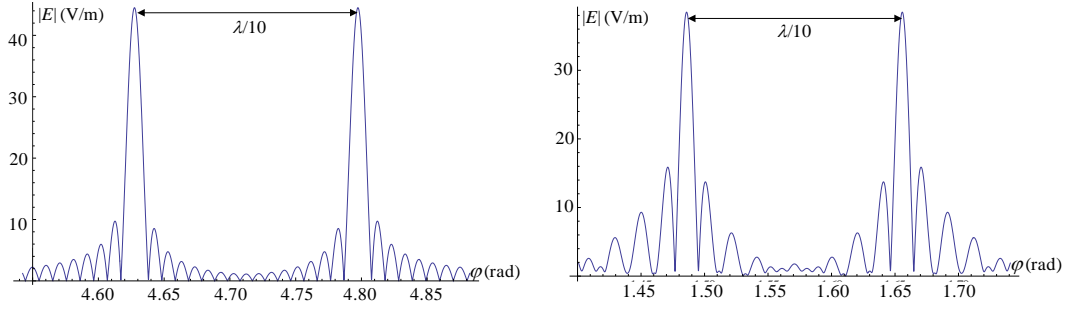

Fig. 6. Modulus of the tangent electric field in the object surface (left) and image surface (right) in the SGW. Between $\theta=\theta_0$ and $\theta=\pi-\theta_0$ there are only incident waves. The separation between the two peak is $\lambda/10$.

## 4. Discussion.

From the point of view of modal theory the radiation emitted by the object surface is expressed as a complete series of incident modes from the object to image surface. Two conditions are necessary to design perfect imaging optical systems:
- The complete set of modes must arrive to the image surface. In the classical optical systems of positive refraction, one part of this set are the evanescent modes, which do not arrive to the image surface, therefore the information contained in these modes is lost. The negative refraction materials permit the recovery of the evanescent modes "amplifying" the value of the exponential decaying fields. In the SGW, the complete set is formed by non-evanescent modes, so the negative refraction material is not necessary to recover the information. For this reason, the system can produce images in the object surface bellow the Abbe limit.
- The coefficients of the series expansion for the fields in the image and object surfaces have to be the same. With this condition, the tangent fields in both surfaces are the same. In an ideal classical system of positive refraction, this condition is fulfilled only for the non-evanescent modes, so the field in the image surface will be always different from the one in the object surface. Since the SGW does not fulfill this condition, the fields in the image and object surfaces differ.

From our point of view the results presented here are the first step for the design of Perfect Imaging systems of better quality using positive refraction. Several subjects are open for future analysis and experiments:
- In the SGW, the refractive index distribution does not depend on $\theta$ (see Eq.(8)). Other structures, as the one shown in Eq. (18) should be analyzed, as well:

$$\left.\begin{array}{l}\varepsilon(r,\theta,\varphi)\\ \mu(r,\theta,\varphi)\end{array}\right\}=\begin{cases}\varepsilon_0\left(\dfrac{R_m}{r}\right)^2 & \mu_0 & \theta_0<\theta<\theta_1\\[2ex] \varepsilon_1\left(\dfrac{R_m}{r}\right)^2 & \mu_1 & \theta_1<\theta<\theta_2\\[2ex] \vdots\\[1ex] \varepsilon_N\left(\dfrac{R_m}{r}\right)^2 & \mu_N & \theta_N<\theta<\pi-\theta_0\end{cases} \qquad (18)$$

- The degree of freedom for the $\varepsilon_i$ and $\mu_i$ permits a better coupling of the modes toward the image surface. The design of these structures is equivalent to the design of the lens coupling wavefronts in the classical optics.
- The Perfect Drain, that is, a perfect receptor at the image surface which does not produce reflection, is another subject that should to be analyzed in details. For one single mode it can be designed a perfect drain material, which eliminates the reflection [17], however, up to our knowledge, there is no any material that can produce null reflection for every mode. We think that the "multilayer" structures, as described in Eq. (18), can be designed also for the receptor to avoid reflection for N modes.
- Finally, there is a very important subject, still not resolved neither with the NRL, the magnification of the lens. The existence of a complete set of non-evanescent modes depends exclusively on the structure. In the case of SGW happens, however, the magnification is only 1. It is important to explore structures where there are non-evanescent modes with higher magnification.

# 3   References.